\documentclass[conference]{IEEEtran}

\usepackage[dvips]{graphicx}
\usepackage{amssymb}

\usepackage[cmex10]{amsmath}
\usepackage[caption=false,font=footnotesize]{subfig}

\newtheorem{theorem}{Theorem}
\newtheorem{remark}{Remark}
\newtheorem{lemma}{Lemma}

\begin{document}

\title{Separate Source-Channel Coding for Broadcasting Correlated Gaussians}

\author{\IEEEauthorblockN{Yang Gao, Ertem Tuncel}
\IEEEauthorblockA{University of California, Riverside, CA\\
Email: yagao@ee.ucr.edu, ertem.tuncel@ucr.edu}}

\maketitle

\begin{abstract}
The problem of broadcasting a pair of correlated Gaussian sources using optimal separate source and channel codes is studied. Considerable performance gains over previously known separate source-channel schemes are observed. Although source-channel separation yields suboptimal performance in general, it is shown that the proposed scheme is very competitive for any bandwidth compression/expansion scenarios. In particular, for a high channel SNR scenario, it can be shown to achieve optimal power-distortion tradeoff.
\end{abstract}

\section{Introduction}
Consider the problem of transmitting two correlated Gaussian sources over a Gaussian broadcast channel with two receivers, each of which desires only to recover one of the sources. 
In \cite{bib: Bross}, it was proven that analog (uncoded) transmission, the simplest possible scheme, is actually optimal when the signal-to-noise ratio (SNR) is below a threshold for the case of matched source and channel bandwidth. 
 To solve the problem for other cases, various hybrid digital/analog (HDA) schemes have been proposed in \cite{bib: Sound, bib: Hamid1, bib: Hamid2}, and \cite{bib: Tian}.
 In fact, the HDA scheme in \cite{bib: Tian} achieves optimal performance for matched bandwidth whenever pure analog transmission does not, thereby leading to a complete characterization of the achievable power-distortion tradeoff.
For the bandwidth-mismatch case, the HDA schemes proposed in \cite{bib: Hamid1} and \cite{bib: Hamid2} comprise of different combinations of previous schemes using either superposition or dirty-paper coding.

In all the aforementioned work, authors also compared achieved performances with that of separate source-channel coding. 
Since the channel is degraded, source coding boils down to sending a ``common'' message to both decoders and a ``refinement'' message to the decoder at the end of the better channel. 
In both of the two source coding schemes proposed in \cite{bib: Sound}, the first source is encoded as the common message, but one scheme encodes (as the refinement message) the second source independently, and the other after {\em de-correlating} it with the first source. In \cite{bib: Tian}, on the other hand, the second source is encoded after it is de-correlated with the {\em reconstruction} of the first source. Although this approach provably yields a better performance than the schemes in \cite{bib: Sound}, it is still not optimal.
In \cite{bib: Tuncel}, it was shown that the optimal rate-distortion (RD) tradeoff in this source coding scenario is in fact achieved by a scheme called successive coding, whereby both common and refinement messages are generated by encoding  both sources jointly, instead of using any kind of de-correlation.
Although successive coding is a special case of successive refinement in its general sense, {\em computation} of the RD tradeoff, even for Gaussians, turned out to be non-trivial. 
A Shannon-type lower bound derived for the problem was rigorously shown to be tight, yielding an analytical characterization of the RD tradeoff.

In this paper, we investigate the performance of separate source and channel coding for any bandwidth compression/expansion ratio. 
As discussed in the previous paragraph, the source coding method to be used for optimal performance is successive coding.
We first show that this separate coding scheme achieves the optimal power-distortion tradeoff when one receiver requires almost lossless recovery, and the other requires a small enough distortion.
Comparing with best-known schemes and outer bounds, we then show that this scheme is competitive in other cases as well.
Our results imply that with a (sometimes marginal) sacrifice of power-distortion performance, we can design separate source and channel codes, and thus enjoy the advantages such as simple extension to different bandwidth compression/expansion ratios.

In Section II, the problem is formally defined. Our main results are proved in Section III and the separate coding scheme is compared with other separation-based schemes and hybrid schemes in Section IV.

\section{Preliminaries}
\begin{figure}[htbp]
\begin{center}
\includegraphics[scale = 0.55]{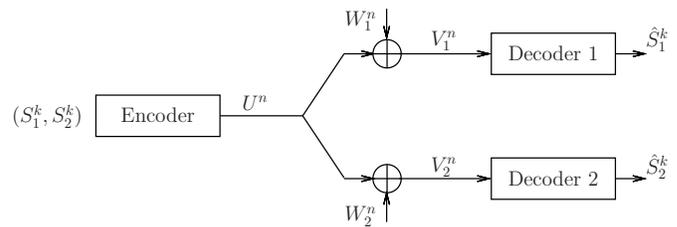}
\caption{System model.}
\label{fig: system}
\end{center}
\end{figure}
As depicted in Fig. \ref{fig: system}, a pair of correlated Gaussian sources $(S_1^k, S_2^k)$ are broadcast to two receivers, and receiver $i$, $i \in\{ 1,2 \}$, is only to reconstruct $S_i^k$.
Without loss of generality, we assume the source sequences are generated in an i.i.d. fashion by $p_{S_1 S_2} = \mathcal{N}(0, \mathbf{C})$, where $$\mathbf{C} = \left[  \begin{array}{cc}  1 & \rho \\ \rho & 1 \end{array}  \right]$$ and $\rho \in [0,1]$. 
The transmitter encodes the source sequences to $U^n$ and thus can be described mathematically as $U^n = \varphi (S_1^k,S_2^k)$. 
We define bandwidth compression/expansion ratio $\kappa = \frac{n}{k}$ with the unit of channel uses per source symbol.

The channel also has an average input power constraint, given by
$$
\frac{1}{n}\sum_{j=1}^{n} E \left[  (U(j))^2 \right] \leq P \; .
$$
At receiver $i$, $U^n$ is corrupted by i.i.d. additive Gaussian noise $W_i^n$, which satisfies $W_i \sim \mathcal{N} (0, N_i) $, where we assume that $N_1 \geq N_2$. 
The channel output $V_i^n$ is then a Gaussian sequence given by $V_i(j) = U(j) + W_i(j)$.
Decoder $1$ reconstructs $S_1^k$ from the channel output $V_1^n$ and can be described as a function $\hat{S}_1^k = \phi_1(V_1^n)$.
Analogously, decoder $2$ computes $\hat{S}_2^k = \phi_2(V_2^n)$.
The reconstruction quality is measured with squared-error distortion, i.e.,
$$
d(s^k,\hat{s}^k) = \frac{1}{k} \sum_{j=1}^k (s_j-\hat{s}_j)^2  \; ,
$$
for any source block $s^k$ and reconstruction block $\hat{s}^k$.
The problem is to find the optimal tradeoff between the channel input power constraint $P$ and the expected distortion pair $(D_1, D_2)$ achieved at the receivers.

In \cite{bib: Sound}, an outer bound to the distortion region is obtained for $\kappa=1$ by assuming full knowledge of $S_1$ at the second (strong) receiver.
In \cite{bib: Hamid2}, that outer bound is extended to bandwidth-mismatched case, in the form of

\setlength{\arraycolsep}{0.14em}
\begin{eqnarray}
\label{eq: ob1}
D_1 &\geq& \left( 1+\frac{\eta P}{\bar{\eta} P + N_1} \right) ^ {-\kappa}  \\
D_2 &\geq& (1-\rho^2 ) \left( 1+\frac{\bar{\eta} P}{N_2} \right) ^ {-\kappa}  \; ,
\label{eq: ob2}  
\end{eqnarray}
where $\eta \in [0,1]$ and $\bar{\eta}=1-\eta$.

Several separation-based schemes have been previously proposed, differing only in their source coding strategy. 
In the first separation-based scheme, termed Scheme A in \cite{bib: Sound}, sources $S_1$ and $S_2$ are encoded as if they are independent, resulting in the distortion region given by
\begin{eqnarray*}
D_1 &\geq& \left( 1+\frac{\eta P}{\bar{\eta} P + N_1} \right) ^ {-\kappa} \\
D_2 &\geq& \left( 1+\frac{\bar{\eta} P}{N_2} \right) ^ {-\kappa}  \; .
\end{eqnarray*}

In Scheme B in \cite{bib: Sound}, the second source is written as $S_2 = \rho S_1 + E$, where $S_1\perp E$, and $S_1$ and $E$ are treated as two new independent sources. Hence we obtain
\begin{eqnarray*}
D_1 &\geq& \left( 1+\frac{\eta P}{\bar{\eta} P + N_1} \right) ^ {-\kappa}  \\
D_2 &\geq& (1-\rho^2)\left( 1+\frac{\bar{\eta} P}{N_2} \right) ^ {-\kappa} + \rho^2 \left( 1+\frac{\eta P}{\bar{\eta} P + N_1} \right) ^ {-\kappa}  \; .
\end{eqnarray*}

In the scheme introduced in \cite{bib: Tian}, which we call Scheme C, $S_1$ is quantized to $\hat{S}_1$ and $S_2$ is then encoded conditioned on $\hat{S}_1$. The resultant distortion region becomes
\begin{eqnarray}
\label{eq: tianD1}
D_1 &\geq& \left( 1+\frac{\eta P}{\bar{\eta} P + N_1} \right) ^ {-\kappa}  \\
D_2 &\geq& \left[ 1-\rho^2  \left( 1-D_1 \right) \right] \left( 1+\frac{\bar{\eta} P}{N_2} \right) ^ {-\kappa}\; .
\label{eq: tianD2}
\end{eqnarray}

Of the three, it is obvious that Scheme C achieves the best performance. However, it is still not optimal as we will show in Section IV. The optimal strategy is in fact what is called successive coding in \cite{bib: Tuncel},  whereby the sources are encoded jointly at both the common and the refinement layers.
The RD tradeoff for successive coding of Gaussian sources with squared-error distortion was given in \cite{bib: Tuncel} parametrically with respect to $\alpha \in [0,1]$ as\footnote{When $D_2>1-\rho^2(1-D_1)$, the optimal strategy degenerates into sending only a common message and estimating $S_2^k$ solely from $\hat{S}_1^k$. So this trivial case is excluded from the discussion in the sequel.}
\begin{eqnarray*}
R_1(\alpha) &=& \frac{1}{2} \log \frac{1-\rho^2}{D_1(1-\nu^2 \delta) - (\rho-\nu \delta)^2}  \\
R_2(\alpha) &=& \left[  \frac{1}{2} \log \frac{1-\nu^2\delta}{D_2} \right]^+\; ,
\end{eqnarray*}
where $\delta = 1-D_1$, $[x]^+ = \max\{x, 0\}$, and
$$
\nu = \left\{ \begin{array}{l} \nu_0 \ ,  {\ \ \ \  \rm if \ \ }  D_2 < 1-\nu_0^2 \delta \\
                                                \nu^* \ ,   {\ \ \ \  \rm if \ \ }  1-\nu_0^2 \delta \leq D_2 < 1-\rho^2\delta
                      \end{array}       \right.
$$
with
$
\nu^* = \sqrt{\frac{1-D_2}{\delta}}
$
, and $\nu_0$ is the unique root of
$$
f_\alpha(\nu) = (1-\alpha)(\rho-\nu\delta)(1-\nu\rho)-\alpha(\nu-\rho)(1-\nu^2 \delta) 
$$
in the interval $[\rho, \min ( \frac{1}{\rho}, \frac{\rho}{\delta} ) ]$.
\section{Main Results}
We first show the RD region of successive coding can be simplified by eliminating both the parameter $\alpha$ and the need to find the roots of the cubic polynomial $f_\alpha(\nu)$.

\begin{lemma}
The achievable source coding rate pair $(R_1, R_2)$, for any distortion pair $(D_1, D_2)$, is given by
\begin{eqnarray}
\label{eq: sourceRate1}
R_1(\nu) &=& \frac{1}{2} \log \frac{1-\rho^2}{D_1(1-\nu^2 \delta) - (\rho-\nu \delta)^2}  \\
R_2(\nu) &=& \left[  \frac{1}{2} \log \frac{1-\nu^2\delta}{D_2} \right]^+\; ,
\label{eq: sourceRate2}
\end{eqnarray}
where $\nu \in \left[ \rho, \min (\frac{1}{\rho}, \frac{\rho}{\delta},\nu^*) \right]$. 
\end{lemma}

The proof is 
deferred to Appendix A.

In separate coding, the region of all achievable $(P, D_1,D_2)$ triplets can be determined using one of two methods. 
The conventional method fixes $P$ and searches for the lower envelope of all $(D_1, D_2)$ whose source rate region intersects with the capacity region given in \cite{bib: Bergmans}.
Alternatively, we can fix $(D_1, D_2)$ and search for the minimum $P$ whose corresponding capacity region intersects with the source rate region given in Lemma $1$.
We find this alternative both more convenient and more meaningful.
More specifically, it is easier to compare schemes based on the minimum power they need to achieve the same distortion pair, and the ratio of minimum powers yields a single number as a quality measure.

To be able to use this alternative, first we need to find out the minimum required power for any given source coding rate pair $(R_1,R_2)$.
\begin{lemma}
For any source coding rate pair $(R_1,R_2)$, the minimal required power is given by
\begin{equation}
P(R_1,R_2) = N_1\left(2^{2 R_1/\kappa}-1\right) + N_2\left(2^{2 R_2/\kappa}-1\right) 2^{2 R_1/\kappa} \; .
\label{eq: mismatchedP}
\end{equation}
\end{lemma}
\begin{IEEEproof} 
For a Gaussian broadcast channel where the better receiver is the second one, $R_1>0$ and $R_2>0$, rates of common and private information, respectively, can be achieved if and only if there exists $0\leq\eta\leq 1$ such that
\begin{eqnarray*}
R_1 & \leq & \frac{\kappa}{2} \log \left(1+\frac{\eta P}{\bar{\eta}P+N_1}\right)\\
R_2 & \leq & \frac{\kappa}{2} \log \left(1+\frac{\bar{\eta} P}{N_2}\right)
\end{eqnarray*}
where $\bar{\eta}=1-\eta$.
This, in turn, implies that $P$ is achievable if and only if there exists $0<\bar{\eta}<2^{- 2 R_1 / \kappa }$ such that 
\[
P \geq \max \left\{\frac{N_1 \left(2^{ 2 R_1/ \kappa }-1\right)}{1-\bar{\eta}2^{ 2 R_1/ \kappa }},\frac{N_2 \left(2^{ 2 R_2 / \kappa }-1\right)}{\bar{\eta}}\right\}  \; .
\]
Since the terms in the maximum exhibit opposite monotonicity with respect to $\bar{\eta}$ with asymptotes at $\bar{\eta}=0$ and $\bar{\eta}=2^{- 2 R_1 / \kappa }$, the minimum power is achieved when the two terms are equal, that is, when
\[
\bar{\eta} = \frac{N_2(2^{ 2 R_2 / \kappa }-1)}{N_1(2^{ 2 R_1 / \kappa }-1) + N_2(2^{ 2 R_2 / \kappa }-1) 2^{ 2 R_1/ \kappa }} \; ,
\]
and has the form in (\ref{eq: mismatchedP}).
\end{IEEEproof}
By substituting  (\ref{eq: sourceRate1}) and (\ref{eq: sourceRate2}) into (\ref{eq: mismatchedP}), we obtain the minimum power required for the separate coding scheme  as a function of $\nu$:
\begin{multline}
P(\nu)  
=  N_1\left( \left[ \frac{1-\rho^2}{D_1(1-\nu^2 \delta) - (\rho-\nu \delta)^2} \right]^ {1/ \kappa }-1\right)  \\
+ N_2\left[ \left( \frac{1-\nu^2 \delta}{D_2} \right) ^{ 1/ \kappa }-1\right] \left[ \frac{1-\rho^2}{D_1(1-\nu^2 \delta) - (\rho-\nu \delta)^2} \right]^ { 1 / \kappa}  \; .
\label{eq: Pnu}
\end{multline}

For bandwidth-matched case, the minimum power of separate coding $P_{\rm sep} = \min_{\nu} P(\nu)$ can actually be found analytically for any $(D_1,D_2)$. We omit the details here.

The following theorem is our first main result.
%
\begin{theorem}
Separate source-channel coding achieves optimal power-distortion tradeoff when $(D_1, D_2)$ satisfies either of the following conditions
\begin{enumerate}
\item $D_1 \rightarrow 0$ and $D_2 \leq 1-\rho^2$ ,
\item $D_2 \rightarrow 0$ and $D_1 \leq 1-\rho^2$ .
\end{enumerate}
\end{theorem}
\begin{IEEEproof}
We first find the minimum power the outer bound (\ref{eq: ob1}) and (\ref{eq: ob2}) requires. Note that when $D_2 > 1-\rho^2$, (\ref{eq: ob2}) will hold for any $\eta \in [0,1]$, and hence the minimum power is obtained solely from (\ref{eq: ob1}), whereas 
when $D_2 \leq 1-\rho^2$, the minimum power satisfies equality in both (\ref{eq: ob1}) and (\ref{eq: ob2}). Combining the two cases, we obtain the concise expression
\begin{equation*}
\begin{split}
&P_{\rm outer\  bound} \\
&=
 N_2D_1^{-1/\kappa}\left[ \left(\frac{D_2}{1-\rho^2} \right)^{-1/\kappa} -1 \right]^+ + N_1(D_1^{-1/\kappa} -1)  \; .
\end{split}
\end{equation*}
On the other hand, from (\ref{eq: Pnu}), we have
\begin{equation*}
P( \rho) 
= N_2D_1^{-1/\kappa} \left[ \left(\frac{D_2}{1-\rho^2 \delta} \right)^{- 1 / \kappa } - 1 \right] + N_1(D_1^{-1 / \kappa } -1)  \; .
\end{equation*}
Since $\delta = 1-D_1$, it is easy to see that when $D_2 \leq 1-\rho^2$ 
$$
\frac{ P(\rho) } {  P_{\rm outer\  bound} }  \rightarrow  1 \ \ \ \ \text{as } D_1 \rightarrow 0 \; ,
$$
and since $\nu = \rho$ is feasible, the minimum power of separate coding satisfies $P_{\rm sep} \leq P(\rho)$. Therefore the performance of separate coding scheme approaches the outer bound, or
$$
\frac{P_{\rm sep} }{ P_{\rm outer\  bound}} \rightarrow 1, \ \ \ \ \text{when } D_2 \leq 1-\rho^2 \text{ and }  D_1 \rightarrow 0 \; .
$$
Similarly, by setting $\nu = \frac{\rho}{\delta}$, we have
\begin{equation*}
\begin{split}
&P\left(\frac{\rho}{\delta}\right) 
=  N_1\left( \left[ \frac{1-\rho^2}{D_1(1-\frac{\rho^2}{ \delta})} \right]^ {1 / \kappa}-1\right)\\
& + N_2\left[ \left( \frac{1-\frac{\rho^2}{ \delta}}{D_2} \right) ^{ 1 / \kappa }-1\right] \left[ \frac{1-\rho^2}{D_1(1-\frac{\rho^2}{ \delta})} \right]^ { 1 / \kappa }  \;,
\end{split}
\end{equation*}
and when $D_2 \rightarrow 0$, $\frac{P(\frac{\rho}{\delta})}{P_{\rm outer\ bound}} \rightarrow1$. Note when $(1-D_1)(1-D_2) \geq \rho^2$, $\min \left( \frac{1}{\rho}, \frac{\rho}{\delta}, \nu^* \right) = \frac{\rho}{\delta}$, which again implies $P_{\rm sep} \leq P(\frac{\rho}{\delta})$, thus proving the second part of the theorem.
\end{IEEEproof}
\begin{remark}
Here we proved that the outer bound is tight in the region of $(1-D_1)(1-D_2) \geq \rho^2$ when either $D_1$ or $D_2$ goes to $0$, and the performance of separate coding approaches the outer bound. 
The condition that either $D_1$ or $D_2$ goes to $0$ translates to infinite channel SNR.
\end{remark}

In fact, as we show in the following theorem, separate coding is approximately optimal for the entire region $(1-D_1)(1-D_2) \geq \rho^2$, in the sense that the power ratio $\frac{P_{\rm sep}}{P_{\rm outer\ bound}}$ can be upper-bounded universally in $(N_1,N_2,D_1,D_2)$.
\begin{theorem}
When $(1-D_1)(1-D_2) \geq \rho^2$,
\begin{multline*}
\frac{P_{\rm sep}}{P_{\rm outer\ bound}} \leq \min \Bigg\{ 1+\frac{(1+\rho)^{1/ \kappa} \left[ (1+\rho^2)^{1/ \kappa}-1 \right] }{(1+\rho)^{1/ \kappa}-(1-\rho)^{1/ \kappa}}, \\
\left( \frac{(1+\rho)^2}{1+2\rho} \right)^{1 / \kappa} + \frac{\left( \frac{(1+\rho)^2}{1+2\rho} \right)^{1 / \kappa} - 1}{(1-\rho^2)^{-1/ \kappa} - 1}\Bigg\}
 \; .
\end{multline*}
\end{theorem}

\begin{IEEEproof}
The first half is true because
\begin{eqnarray*}
& &\frac{P_{\rm sep}}{P_{\rm outer\ bound}} \stackrel{(a)}{\leq} \frac{P(\rho)}{P_{\rm outer\ bound}} \\
& = & \frac{\frac{N_2}{N_1}\left[ (1-\rho^2 \delta)^{1/ \kappa}  - D_2^{1/ \kappa} \right] + D_2^{1/ \kappa} (1- D_1^{1/ \kappa})}{\frac{N_2}{N_1}\left[ (1-\rho^2)^{1/ \kappa}  - D_2^{1/ \kappa} \right] + D_2^{1/ \kappa} (1- D_1^{1/ \kappa})}  \\
& \stackrel{(b)}{\leq} & \frac{(1-\rho^2 \delta)^{1/ \kappa} - (D_1 D_2)^{1/ \kappa} }{(1-\rho^2)^{1/ \kappa} - (D_1 D_2)^{1/ \kappa}}  \\
& \stackrel{(c)}{\leq} & \frac{(1-\rho^2 \delta)^{1/ \kappa} - (1-\rho)^{2/ \kappa}}{(1-\rho^2)^{1/ \kappa} - (1-\rho)^{2/ \kappa}} \\
& \stackrel{(d)}{\leq} & 1+\frac{(1+\rho)^{1/ \kappa} \left[ (1+\rho^2)^{1/ \kappa}-1 \right] }{(1+\rho)^{1/ \kappa}-(1-\rho)^{1/ \kappa}}\; , 
\end{eqnarray*}
where $(a)$ follows since $\nu = \rho$ is feasible, $(b)$ by $\frac{N_2}{N_1} \leq 1$, $(c)$ follows since $D_1D_2 \leq (1-\rho)^2$, and $(d)$ since $\delta \geq \rho^2$.

By relaxing $P_{\rm sep}$ to $P\left(\frac{\rho}{\sqrt{\delta}} \right)$, the second half of the bound can be obtained in a similar way.
The detailed proof is omitted here due to lack of space.
\end{IEEEproof}

\section{Performance comparison}
%
\subsection{Separation-based schemes}
%
\begin{figure}[!ht]
\centering
\subfloat[Power difference between separate coding and the outer bound.]{\label{Tsubfig1}\includegraphics[width = 0.5\textwidth]{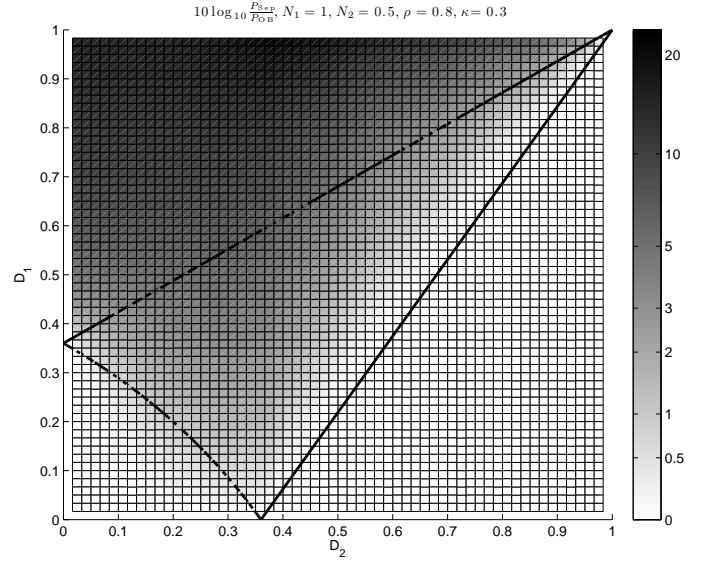}} \\
\subfloat[Power difference between Scheme C and optimal separate coding.]{\label{Tsubfig2}\includegraphics[width = 0.5\textwidth]{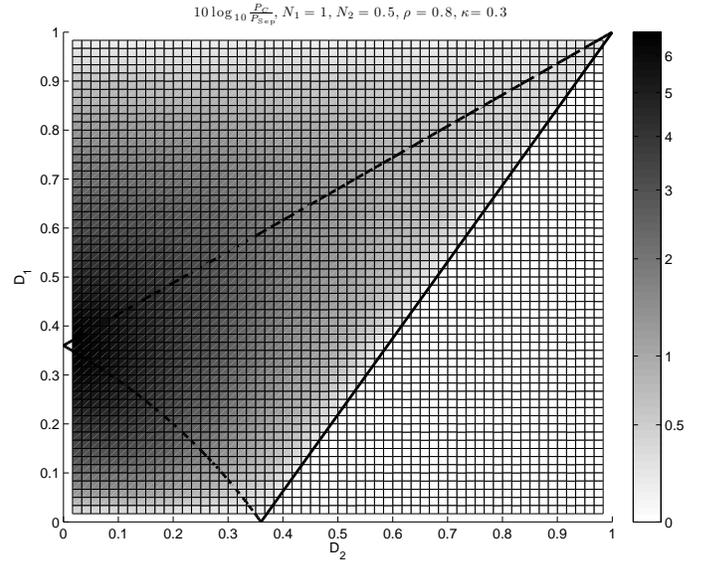}} 
\caption{Comparison between the power of outer bound, Scheme C and optimal separate coding. $\rho = 0.8$, $N_1 =  1$, $N_2 = 0.5$ and $\kappa=0.3$.}
\label{fig: tian}
\end{figure}
As illustrated in \cite{bib: Tian} for $\kappa = 1$, the outer bound in (\ref{eq: ob1}) and   (\ref{eq: ob2}) is not always tight. Nevertheless, we can still compare $P_{\rm sep}$ and $P_{\rm outer\ bound}$ for any $(D_1,D_2)$, which provides an upper bound to the ratio of the minimum separate and joint coding power levels.
To show the optimality of our separate coding scheme, we also compare our scheme with Scheme C, which provides the best performance among the three separation-based schemes mentioned earlier.
The minimum required power of Scheme C can be obtained from (\ref{eq: tianD1}) and (\ref{eq: tianD2}) as
\begin{equation*}
P_C =  N_2D_1^{-1/\kappa}\left[ \left(\frac{D_2}{1-\rho^2 \delta} \right)^{-1/\kappa} -1 \right] + N_1(D_1^{-1/\kappa} -1)  \; .
\end{equation*}
Note that $P_C = P(\rho)$ in the non-trivial distortion regions, so it is immediately clear that the successive coding scheme outperforms Scheme C.

As an example with bandwidth compression, we show the power ratio between our separate coding scheme and the outer bound in Figure \ref{fig: tian}\subref{Tsubfig1}, and that between Scheme C and our separate coding scheme in Figure \ref{fig: tian}\subref{Tsubfig2}, both in dB.
For reference, the black curves illustrate the different distortion regions for a related problem in \cite{bib: Xiao}, where only one receiver is in presence and interested in reconstructing both sources. 
The lower left corner region is actually $(1-D_1)(1-D_2) \geq \rho^2$, where, in general, small dB differences are observed, as implied by the two theorems above.
As can be seen from the figure, even for highly correlated sources, the optimum separate coding scheme does not require too much extra power in most of the $(D_1,D_2)$ plane.
Again, since the outer bound is not always tight, the large power difference in some regions may be dramatically reduced when the outer bound is replaced by the optimum performance.
For smaller $\rho$ values, we observe that the power difference is very small in the entire plane, as an example illustrates in the next section.
This is a natural result because separate coding is optimal for independent sources and small $\rho$ value means the sources are not highly dependent.

There is also noticeable power difference between our scheme and Scheme C, and we numerically observe large dB values near the point $D_1=1-\rho^2$ and $D_2=0$, for which we can obtain analytically the power ratio.

\begin{theorem}
When $D_1 = 1-\rho^2$ and $D_2 \rightarrow 0$, 
$$
\frac{P_C}{P_{\rm sep}} \rightarrow  (1+\rho^2)^{1/ \kappa} \; .
$$
\end{theorem}
The proof uses the fact that the power of our separate coding scheme goes to that of the outer bound when approaching this point.

When the two receivers have the same noise level, i.e., $N_1 = N_2$, it can be shown that our separate coding scheme achieves the corresponding rate-distortion function $R(D_1,D_2)$ in \cite{bib: Xiao}, whereas none of the three separate coding schemes mentioned earlier has the same performance.
\subsection{Hybrid Digital/Analog (HDA) schemes}
%
\begin{figure}[!ht]
\centering
\subfloat[$\rho=0.2$]{\label{subfig1}\includegraphics[width = 0.5\textwidth]{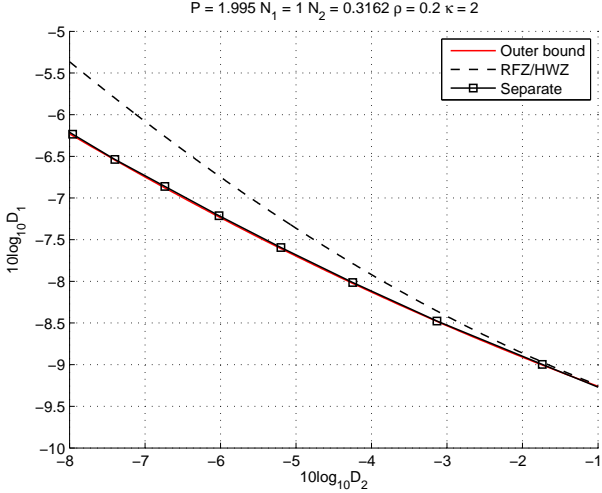}} \\
\subfloat[$\rho=0.8$]{\label{subfig2}\includegraphics[width = 0.5\textwidth]{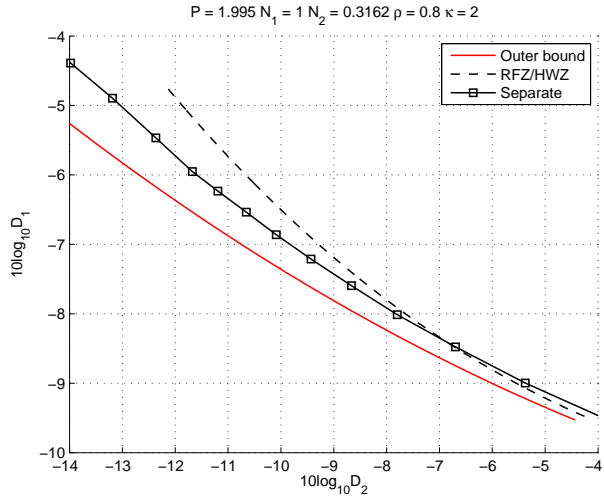}} 
\caption{Comparison between outer bound, RFZ scheme in \cite{bib: Hamid2} and separate coding. $P=3dB = 1.995$, $N_1 = 0dB =1$, $N_2 = -5dB = 0.3162$ and $\kappa=2$.}
\label{fig: hamid}
\end{figure}
In \cite{bib: Hamid1} and \cite{bib: Hamid2},  a group of hybrid digital/analog (HDA) schemes were proposed for  bandwidth-mismatched case, where analog, digital, and hybrid schemes are layered with superposition or dirty-paper-coding.
The achievable distortion region can be found by varying power allocation and scaling coefficients.
In \cite{bib: Hamid2}, an HDA scheme from \cite{bib: Reznic} for broadcasting a common source with bandwidth expansion was adapted for the problem of broadcasting correlated sources and is termed the RFZ scheme.
In addition, a scheme, termed the HWZ scheme, containing an analog layer and two digital layers each with a Wyner-Ziv coder and a channel coder, was also proposed.
It is argued  by an example in \cite{bib: Hamid2} that the HWZ scheme performs similar to the RFZ scheme.
Here we compare our separate coding scheme with the outer bound and the RFZ/HWZ scheme in Figure \ref{fig: hamid}.\footnote{The performance of RFZ is plotted to represent both schemes as their curves almost coincide at least for this set of parameters in \cite{bib: Hamid2}.}
For this comparison, we revert to the more familiar $(D_1,D_2)$ plot for the exact same $(P,\rho,\kappa,N_1,N_2)$ as those used in the examples in \cite{bib: Hamid2}.

As seen in Figure \ref{fig: hamid}\subref{subfig1}, when $\rho$ is small, the separate coding scheme almost coincides with the outer bound and outperforms RFZ/HWZ schemes.
When the sources are highly correlated as in Figure \ref{fig: hamid}\subref{subfig2}, the separate coding scheme is still better than the RFZ/HWZ schemes when $D_2$ is lower than a certain value, and also provides competitive performance when it is higher.
We observed similar performance behavior when we compared the separate coding scheme to the layered schemes in \cite{bib: Hamid1} for bandwidth compression. 

\section{Conclusion}
The performance of optimum separate source-channel coding scheme for broadcasting two correlated Gaussians is analyzed. 
The minimum power required for a given distortion pair is used as a tool to compare performances of different schemes. 
It is illustrated that this separate coding scheme outperforms other known separate schemes, and is competitive in general in the sense that its minimum required power is close to the power implied by the outer bound.   
Also, in a certain ``low distortion'' region, the power difference is analytically bounded.
In fact, within this region, in the extreme cases of almost lossless reconstruction of either source, the separate scheme is provably optimal.

\appendices
\section{Proof of Lemma $1$}
The cubic function in \cite{bib: Tuncel} is
$$
f_\eta (\nu) = (1-\eta)(\rho - \nu \delta) (1- \nu \rho) -\eta (\nu -\rho) (1 - \nu^2 \delta)
$$
and when $f_\eta (\nu) = 0$, it can be re-written as
$$
\eta = \frac{(\rho - \nu \delta) (1 - \nu \rho )}{ \nu [1-\rho^2 - \delta (1-2\nu \rho +\nu^2)]}  \; .
$$
It will be shown that varying $\eta$ in $[0,1]$ is equivalent with varying $\nu$ in $\left[ \rho, \min (\frac{1}{\rho}, \frac{\rho}{\delta}, \sqrt{\frac{1-D_2}{\delta}}) \right]$, by showing $\eta$ is a monotonically decreasing function of $\nu$.

When $\delta > \rho^2$, $\rho < \frac{\rho}{\delta} < \frac{1}{\rho}$ and also note $\eta = 1 {\rm \ and\ } 0$  when $\nu = \rho {\rm \ and\ } \frac{\rho}{\delta}$, respectively. We examine
$$
h(\nu) = \frac{\rho-\nu\delta}{1-\rho^2 - \delta (1-2\nu \rho +\nu^2)}
$$
instead of $\eta$, and
$$
\frac{d h}{d \nu} \propto -1-\rho^2+\delta+2\nu \rho -\nu^2 \delta  \; .
$$
The right hand side is a quadratic function of $\nu$ centered at $\nu = \frac{\rho}{\delta}$ and the maximum value is $-(1- \delta)(1-\frac{\rho^2}{\delta}) \leq 0$. (When $\nu = \rho$, the function value is $-(1-\rho^2)(1-\delta)$.)

Similarly, when $\delta < \rho^2$, we have $\rho < \frac{1}{\rho} < \frac{\rho}{\delta}$ and in this case, $\eta = 1 {\rm \ and\ } 0$  when $\nu = \rho {\rm \ and\ } \frac{1}{\rho}$. We examine
$$
h(\nu) = \frac{1 - \nu \rho }{1-\rho^2 - \delta (1-2\nu \rho +\nu^2)}  \; ,
$$
and thus have
$$
\frac{d h}{d \nu} \propto -\rho(1-\rho^2)-\rho \delta+2\nu \delta -\nu^2 \rho \delta  \; .
$$
The right hand side is centered at $\nu = \frac{1}{\rho}$ and the maximum value is $-(1-\rho^2)(\rho-\frac{\delta}{\rho}) \leq 0$. 
\hspace*{\fill}$\blacksquare$\par

\end{document}